\documentclass[12pt,english,floatfix,superscriptaddress,aps,prd,preprint,showkeys]{revtex4}
\usepackage[utf8]{inputenc}
\usepackage{amsmath}
\usepackage{amssymb}
\usepackage{amsbsy}
\usepackage{amsfonts}
\usepackage{amsopn}
\usepackage{amstext}
\usepackage{graphicx}
\usepackage{amssymb}
\usepackage{amsfonts}
\usepackage{amsmath}
\usepackage{amsmath,amsthm,amsfonts,amssymb}
\usepackage[mathcal]{eucal}
\usepackage{mathrsfs}
\usepackage{graphicx}
\usepackage[english]{babel}
\usepackage{color}
\usepackage{esint}
\usepackage[dvips]{epsfig}
\usepackage[dvips]{graphicx}
\usepackage{float}
\usepackage{units}
\usepackage{textcomp}

\usepackage{hyperref}
\usepackage{slashed}

\newcommand{\ie}{\begin{equation}}
\newcommand{\fe}{\end{equation}}
\newcommand{\se}{\begin{eqnarray}}
\newcommand{\ff}{\end{eqnarray}}

\begin{document}

\title{Thermodynamic properties of neutral Dirac particles in the presence of an electromagnetic field}

\author{R. R. S. Oliveira}
\email{rubensrso@fisica.ufc.br}
\affiliation{Universidade Federal do Cear\'a (UFC), Departamento de F\'isica,\\ Campus do Pici, Fortaleza - CE, C.P. 6030, 60455-760 - Brazil.}


\author{A. A. Ara\'ujo Filho}
\email{dilto@fisica.ufc.br}
\affiliation{Universidade Federal do Cear\'a (UFC), Departamento de F\'isica,\\ Campus do Pici, Fortaleza - CE, C.P. 6030, 60455-760 - Brazil.}


\date{\today}

\begin{abstract}

In this paper, we investigate the thermodynamic properties of a set of neutral Dirac particles in the presence of an electromagnetic field in contact with a heat bath for the relativistic and non-relativistic cases. In order to perform the calculations, the high-temperature limit is considered and the \textit{Euler-MacLaurin} formula is taking into account. Next, we explicitly determine the behavior of the main thermodynamic functions of the canonical ensemble: the Helmholtz free energy, the mean energy, the entropy, and the heat capacity. As a result, we verified that the mean energy and the heat capacity for the relativistic case are two times the values of the non-relativistic case, thus, satisfying the so-called \textit{Dulong-Petit} law. In addition, we also verified that the Helmholtz free energy and the entropy in both cases increase as a function of the electric field. Finally, we note that there exists no influence on the thermodynamic functions due to the magnetic field. 
\end{abstract}


\maketitle

\section{Introduction}

The study of the physical properties of materials, focusing on its thermodynamic properties, is of great interest in condensed matter physics, solid-state physics, and materials science \cite{Gaskell,DeHoff,Tester,Dolling,Muhlschlegel,Coleman,Eckert,Foiles,Anthony,Wang}. Indeed, the efforts spending to obtain the knowledge of thermodynamic properties are justified by both practical needs and fundamental science \cite{Balandin}. Some examples with practical relevance are shown in Refs. \cite{Balandin,Mounet,Shahil,Pop,Alofi,Che,Ruoff}, where are investigated the thermodynamic properties of graphene, diamond, graphite, carbon nanotubes, nanostructured carbon materials. Moreover, this study also has a particular relevance from a theoretical viewpoint, mainly when the physical system is either relativistic \cite{Boumali,Hassanabadi,Pacheco,Wang2015,Arda} or non-relativistic quantum \cite{Groote,Dong,Nammas,Oyewumi,Arda2016}. In addition, it is worth mentioning that recently the thermodynamic properties of an Aharonov-Bohm quantum ring were performed in the high and low-energy regime \cite{Oli}.

On the other hand, the Dirac equation (DE) for spin-1/2 neutral particles with magnetic dipole moment (MDM) in the presence of electric and magnetic fields also has many remarkable applications. For instance, it is applied to problems involving the Aharonov-Casher effect \cite{Hagen,Mirza,Li}, cosmic string spacetime \cite{Bakke,Oliveira}, electric dipole moment \cite{Silenko}, electromagnetic waves \cite{Khalilov}, scattering \cite{Lin}, spin effects \cite{Azevedo}, noninertial effects \cite{Bakke2010}, geometric phases in the presence of torsion, topological defect \cite{Bakke2008}, and quantum dots \cite{Bakke2012}. Besides that, it is worth mentioning that there exists another physical approach which has received much attention over the years, namely, the so-called Dirac oscillator. It was first introduced in the context of DE for particles with MDM interacting with external electric fields \cite{Moshinsky,Martinez,O}.
Moreover, recently the thermodynamic properties of spinless neutral particles with MDM in the presence of topological defects were calculated in the low-energy regime \cite{H}.

The present paper has the purpose of determining the thermodynamic properties of neutral Dirac particles with MDM in the presence of an external electromagnetic field. In particular, we determine the thermodynamic properties for both relativistic and non-relativistic cases. Furthermore, we focus on the high-temperature regime and we regard a set of indistinguishable noninteracting $N$-particles in contact with a thermal reservoir. In order to perform the calculations, we use the \textit{Euler-MacLaurin} sum formula to construct the canonical partition function of the system. Next, we explicitly determine the thermodynamic quantities of interest, namely, Helmholtz free energy, the mean energy, the entropy, and the heat capacity for relativistic and non-relativistic cases.

This paper is organized as follows. In Section \ref{sec2}, we determine explicitly the thermodynamic properties of a set of neutral Dirac particles, such as the Helmholtz free energy, the mean energy, the entropy and the heat capacity. In Section \ref{sec3}, we present the results and discussions about the behavior of the thermodynamic functions in the high-temperature regime. In Section \ref{conclusion}, we finish our work with the conclusions and some final remarks.

\section{Thermodynamic properties of Dirac fermions\label{sec2}}

For both relativistic and non-relativistic cases, in this section, we calculate the thermodynamic properties of a set of Dirac $N$-particles in contact with a thermal reservoir at constant temperature $T$. These properties are given by the following thermodynamic quantities: the Helmholtz free energy, the mean energy, the entropy and the heat capacity. In this sense, our discussion begins initially with the discrete energy spectrum for relativistic Dirac particles with MDM $\mu$ in the presence of an external electromagnetic field, given by Ref.\cite{Rubens}, namely
\ie E^\sigma_{n_s,m_j}=\mu B+\sigma m_0 c^2\sqrt{1+\frac{\hbar\omega_{AC}}{m_0 c^2}[2n_s+\vert m_j\vert+\delta m_j]}, \ \ (n=0, 1, 2,\ldots; m_j=\pm 1,\pm 2,\pm 3,\ldots),
\label{energy}\fe
where $n_s=n+\frac{1-s}{2}$, being $n$ a quantum number and the parameter $s=\pm 1$ describes the two components of the Dirac spinor, $\sigma=+1$ corresponds to the positive energy states (particle with $s=+1$), $\sigma=-1$ corresponds to the negative energy
states (antiparticle with $s=-1$), the parameter $\delta=\mp 1$ describes a negatively ($\delta=-1$) or positively ($\delta=+1$) charged cylinder, $B$ is the strength of the uniform magnetic field and $\omega_{AC}$ is the cyclotron frequency. 

In addition, the non-relativistic energy spectrum is obtained considering that greater part of the total energy of the system lies in the rest energy of the particle \cite{Oliveira}, i.e., $E=m_0 c^2+\varepsilon$, where $m_0 c^2\gg\varepsilon$ and $m_0 c^2\gg\mu B$. So, applying this prescription in \eqref{energy}, we obtain the following energy spectrum for a spinless non-relativistic particle in the presence of an external electromagnetic field
\ie\varepsilon_{n,m_j}=\mu B+\frac{1}{2}\hbar\omega_{AC}[2n+\vert m_j\vert+\delta m_j]>0, \ \ (s=+1),
\label{energy2}\fe

In particular, for $\delta m_j>0$, the spectra \eqref{energy} and \eqref{energy2} present finite degeneracy, while for $\delta m_j<0$, presents an infinite degeneracy \cite{Rubens}. In addition, since it has a significant influence in the thermodynamic properties of a physical system \cite{Pacheco,Arda2016,Greiner}, we will work on the spectra that represent only finite degeneracy.

\subsection{The relativistic case\label{subsec1}}

In this case, let us start our discussion with the fundamental object of the statistical mechanics for the canonical ensemble, the so-called partition function $Z$, which is defined as the sum of all possible quantum states of the system \cite{Greiner}. Explicitly, the one-particle partition function is written by following expression \cite{Pacheco,Greiner}
\ie Z(T,V,1)=\sum_{k=1}^\infty\Omega(E_k)e^{-\beta E_k},
\label{partition}\fe
where $\beta=\frac{1}{k_B T}$, being $k_B$ the Boltzmann constant, $T$ is the thermodynamic equilibrium temperature and $\Omega(E_n)$ is the degree of degeneracy for the energy level $E_k$ defined as
\ie E_k=\mu B+m_0 c^2\sqrt{1+2\xi k},
\label{energy3}\fe
where $k=n+\vert m_j\vert\geq 1$, being $k$ a new quantum number and $\xi=\frac{\hbar\omega_{AC}}{m_0 c^2}$ is a dimensionless constant parameter. In special, to obtain the spectrum \eqref{energy3}, we consider a particle (with $E>0$ and $\sigma=s=+1$) interacting with a negatively ($\delta=-1$) charged cylinder. To determine $\Omega(E_k)$, let us note that for each quantum level with ($n,\vert m_j\vert$) there are $2\vert m_j\vert+1$ degenerate states differing with values of orbit magnetic quantum number $m_l=\pm\frac{1}{2},\pm\frac{3}{2},\ldots$ \cite{Pacheco,Arda2016,Rubens}. For a given $k$ the total degree of degeneracy is obtained as
\ie \Omega(E_k)=\sum_{\vert m_j\vert=1}^k(2\vert m_j\vert+1)=k(k+2),
\label{degeneracy}\fe
consequently, the partition function \eqref{partition} becomes
\ie Z(T,V,1)=\sum_{k=1}^\infty k(k+2)e^{-(a+b\sqrt{1+2\xi k})},
\label{partition1}\fe
where $a=\beta\mu B$ and $b=\beta m_0 c^2$. 

Before proceeding further, it is recommended to analyze the convergence of the partition function \cite{Pacheco}. So, the function $f(x)=x(x+2)e^{-(a+b\sqrt{1+2\xi x})}$ is a monotonically decreasing function and the corresponding integral
\ie\
\begin{split}
I(a,b)=\int_{1}^{\infty}x(x+2)e^{-(a+b\sqrt{1+2\xi x})} dx=\left[\frac{3}{b\xi}\sqrt{1+2\xi}+\frac{1}{b^2\xi^2}(11\xi+4)\right]e^{-(a+b\sqrt{1+2\xi})}       
& \\ +\left[\frac{1}{b^3\xi^3}\sqrt{1+2\xi}(16\xi+2)+\frac{1}{b^4\xi^3}(36\xi+12)+\frac{3}{b^5\xi^3}\sqrt{1+2\xi}+\frac{30}{b^6\xi^3}\right]e^{-(a+b\sqrt{1+2\xi})},
\end{split}
\label{integral}\fe
is convergent. Thus from the theorems of convergent series, this implies that the partition function also is convergent.

However, the partition function \eqref{partition1} cannot be exactly calculated in a closed form, rather for high temperatures ($T\to\infty$), as well as for low temperatures ($T\to 0$), we can obtain reseanable approximations \cite{Greiner}. In particular, a sytemmatic expansion of \eqref{partition1} for large $T$ is possible with the use of the \textit{Euler-MacLaurin} sum formula, with serves to calculate the integrals numerically. The \textit{Euler-MacLaurin} sum formula is given by \cite{Greiner}
\ie Z(T,V,1)=\sum_{k=1}^{\infty}f(k)=\frac{1}{2}f(1)+\int_{1}^{\infty}f(x) dx-\sum_{p=1}^{\infty}\frac{1}{(2p)!}B_{2p}f^{(2p-1)}(1),
\label{partition2}\fe
or simply as
\ie Z(T,V,1)=\sum_{k=1}^{\infty}f(k)=\frac{1}{2}f(1)+\int_{1}^{\infty}f(x) dx-\frac{1}{12}f'(1)+\frac{1}{720}f'''(1)-\ldots+,
\label{partition3}\fe
where $B_{2p}$ are the Bernoulli numbers.

So, the partition function \eqref{partition3} is given by
\ie\
\begin{split}
Z(T,V,1)=\left[\frac{7}{6}+\frac{3}{b\xi}\sqrt{1+2\xi}+\frac{1}{b^2\xi^2}(11\xi+4)+\frac{1}{b^3\xi^3}\sqrt{1+2\xi}(16\xi+2)\right]e^{-(a+b\sqrt{1+2\xi})} \ \ \      
& \\ +\left[\frac{1}{b^4\xi^3}(36\xi+12)+\frac{3}{b^5\xi^3}\sqrt{1+2\xi}+\frac{30}{b^6\xi^3}+\frac{1}{4}\frac{b\xi}{\sqrt{1+2\xi}}+\frac{12}{720}\frac{b^2\xi^2}{(1+2\xi)}\right]e^{-(a+b\sqrt{1+2\xi})} \ \ \
& \\ -\frac{1}{720}\left[\frac{2b\xi}{\sqrt{1+2\xi}}+\frac{b^3\xi^3}{(1+2\xi)^{3/2}}+\frac{3b^2\xi^3}{(1+2\xi)^2}+\frac{3b\xi^3}{(1+2\xi)^{5/2}}-\frac{b\xi^2}{(1+2\xi)^{3/2}}\right]e^{-(a+b\sqrt{1+2\xi})}+\mathcal{O}(b^4)\
\end{split}.
\label{partition4}\fe

Considering the high temperatures regime where $a\ll 1$ and $b\ll 1$, the partition function \eqref{partition4} becomes
\ie Z(T,V,1)\simeq\left(\frac{30}{b^6\xi^3}\right),
\label{partition5}\fe
consequently, for $N$-particles we have
\ie Z(T,V,N)\simeq\left(\frac{30}{b^6\xi^3}\right)^N.
\label{partition6}\fe

Now, let us concentrate in the  main thermodynamic quantities. In special, the quantities of our interest are the Helmholtz free energy $F$, the mean energy $U$, the entropy $S$ and the heat capacity $C_V$ \cite{Pacheco,Greiner}, which are defined as follows
\ie F=-\frac{1}{\beta}\ \mathsf{ln}\ Z, \ \ U=-\frac{\partial}{\partial\beta}\ \mathsf{ln}\ Z, \ \ S=k_B\beta^2\frac{\partial}{\partial\beta}F, \ \ C_V=-k_B\beta^2\frac{\partial}{\partial\beta}U.
\label{properties}\fe

Therefore, using partition function \eqref{partition6}, the Helmholtz free energy, the mean energy, the entropy and the heat capacity for the relativistic case are written as
\ie \bar{F}\simeq -\tau\ \mathsf{ln}\left(\frac{30\tau^6}{\xi^3}\right), \ \ \bar{U}\simeq 6\tau, \ \ \bar{S}\simeq\left[6+\mathsf{ln}\left(\frac{30\tau^6}{\xi^3}\right)\right], \ \ \bar{C}_V\simeq 6,
\label{properties1}\fe
where $\bar{F}=\frac{F}{N m_0 c^2}$, $\bar{U}=\frac{U}{N m_0 c^2}$, $\bar{S}=\frac{S}{N k_B}$, $\bar{C}_V=\frac{C_V}{N k_B}$ and $\tau$ is a parameter given by $\tau=\frac{k_B T}{m_0 c^2}=\frac{T}{T_0}$, as $T_0=\frac{m_0 c^2}{k_B}\simeq 5.93 \times 10^9$ K stands for the characteristic temperature of the system and is analogous to the so-called Debye temperature defined in solid state physics \cite{Pacheco}. From a dimensional point of view, all the quantities in \eqref{properties1} are coherent.

\subsection{The non-relativistic case\label{subsec2}}

Now, let us take into account the thermodynamic properties for the non-relativistic case where most phenomena of condensed matter physics and solid state occur. So, the one-particle canonical partition function is written by the following expression \cite{Pacheco,Greiner}
\ie Z(T,V,1)=\sum_{k=1}^\infty\Omega(\varepsilon_k)e^{-\beta\varepsilon_k},
\label{partition7}\fe
where $\Omega(\varepsilon_k)$ is the degree of degeneracy for the energy level $\varepsilon_k$ defined as
\ie\varepsilon_k=\mu B+\bar{\xi}k, \ \
\label{energy4}\fe
where $k=n+\vert m_j\vert\geq 1$ and $\bar{\xi}=\hbar\omega_{AC}$. In special, to obtain the spectrum \eqref{energy4} we again consider a particle ($\sigma=s=+1$) interacting with a negatively ($\delta=-1$) charged cylinder. To determine $\Omega(\varepsilon_k)$, let us note that for each quantum level with ($n,\vert m_j\vert$) there are $2\vert m_j\vert+1$ degenerate states differing with values of orbit magnetic quantum number $m_l=\pm\frac{1}{2},\pm\frac{3}{2},\ldots$ \cite{Pacheco,Arda2016,Rubens}. For a given $k$, the total degree of degeneracy is obtained as
\ie \Omega(\varepsilon_k)=\sum_{\vert m_j\vert=1}^k(2\vert m_j\vert+1)=k(k+2),
\label{degeneracy1}\fe
and consequently, partition function \eqref{partition7} becomes
\ie Z(T,V,1)=\sum_{k=1}^\infty k(k+2)e^{-(a+\bar{b}k)},
\label{partition8}\fe
where $a=\beta\mu B$ and $\bar{b}=\beta\bar{\xi}$. 

Analogously  to the relativistic case,  it is also reasonable to analyze the convergence of the non-relativistic partition function \cite{Pacheco}. In this way, the function $f(x)=x(x+2)e^{-(a+\bar{b}x)}$ is a monotonically decreasing function and the corresponding integral
\ie \bar{I}(a,\bar{b})=\int_{1}^{\infty}x(x+2)e^{-(a+\bar{b}k)} dx=\left[\frac{3}{\bar{b}}+\frac{4}{\bar{b}^2}+\frac{2}{\bar{b}^3}\right]e^{-(a+\bar{b})}
\label{integral1}\fe
is also convergent.

Again, using the \textit{Euler-MacLaurin} sum formula given in \eqref{partition3}, the partition function \eqref{partition8} is rewritten as
\ie Z(T,V,1)=\left[\frac{7}{6}+\frac{3}{\bar{b}}+\frac{4}{\bar{b}^2}+\frac{2}{\bar{b}^3}+\frac{29}{120}\bar{b}+\frac{1}{60}\bar{b}^2-\frac{1}{240}\bar{b}^3\right]e^{-(a+\bar{b}k)}+\mathcal{O}(\bar{b}^4).
\label{partition9}\fe

Considering the high temperatures regime where $a\ll 1$ and $\bar{b}\ll 1$, we have
\ie Z(T,V,1)\simeq\left(\frac{2}{\bar{b}^3}\right),
\label{partition10}\fe
and consequently, the total partition function for a set of $N$-particles is given by
\ie Z(T,V,N)\simeq\left(\frac{2}{\bar{b}^3}\right)^N.
\label{partition11}\fe

Now, let us concentrate our efforts in the  main thermodynamic quantities. Therefore, using relations \eqref{properties} and the partition function \eqref{partition11}, the Helmholtz free energy, the mean energy, the entropy and the heat capacity for the non-relativistic case are written as
\ie \bar{f}\simeq -k_B T\ \mathsf{ln}\left(\frac{2k_B^3 T^3}{\bar{\xi}^3}\right), \ \ \bar{u}\simeq 3k_B T, \ \ \bar{s}\simeq\left[3+\mathsf{ln}\left(\frac{2k_B^3 T^3}{\bar{\xi}^3}\right)\right], \ \ \bar{c}_V\simeq 3,
\label{properties2}\fe
where $\bar{f}=\frac{F}{N}$, $\bar{u}=\frac{U}{N}$, $\bar{s}=\frac{S}{N k_B}$, $\bar{c}_V=\frac{C_V}{N k_B}$.

\section{Results and discussions\label{sec3}}

Initially, we display our results on the calculation of the thermodynamic functions for the relativistic case. Since we obtained a degenerate spectrum, we use the \textit{Euler-MacLaurin} sum formula and consider the high temperatures regime to evaluate numerically the partition function. Nevertheless, through some steps analogous to the relativistic case, we provide the calculations of the same thermodynamic functions for the non-relativistic case as well.

\begin{figure}[ht]
\centering
\includegraphics[scale=0.6]{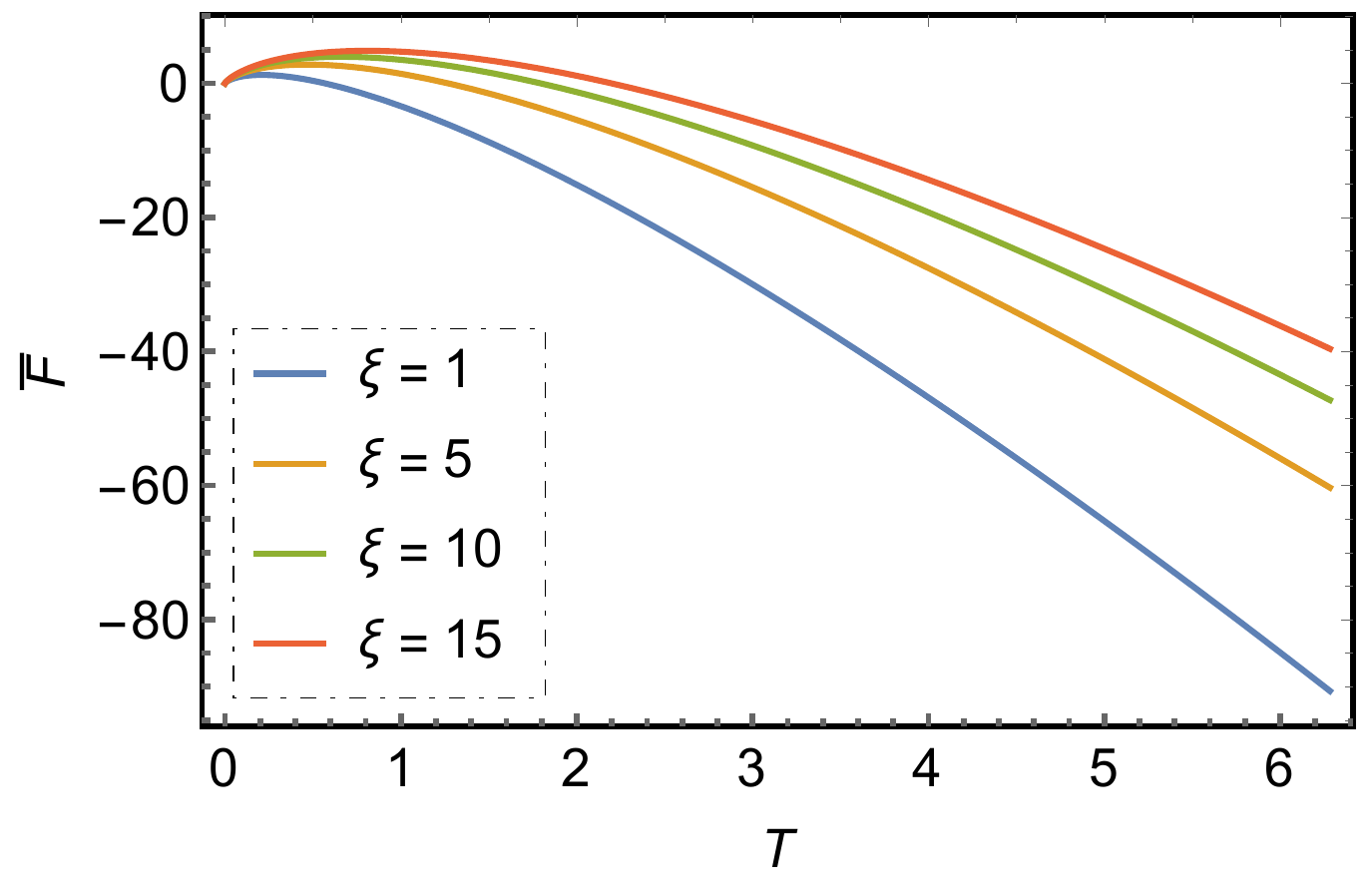}
\caption{Regarding the relativistic case, the graphic exhibits the behavior of the Helmholtz free energy $\bar{F}$ of neutral Dirac particles in the presence of electromagnetic field for different values of $\xi$ when the temperature $T$ increases.}
\label{1}
\end{figure}

\begin{figure}[ht]
\centering
\includegraphics[scale=0.415]{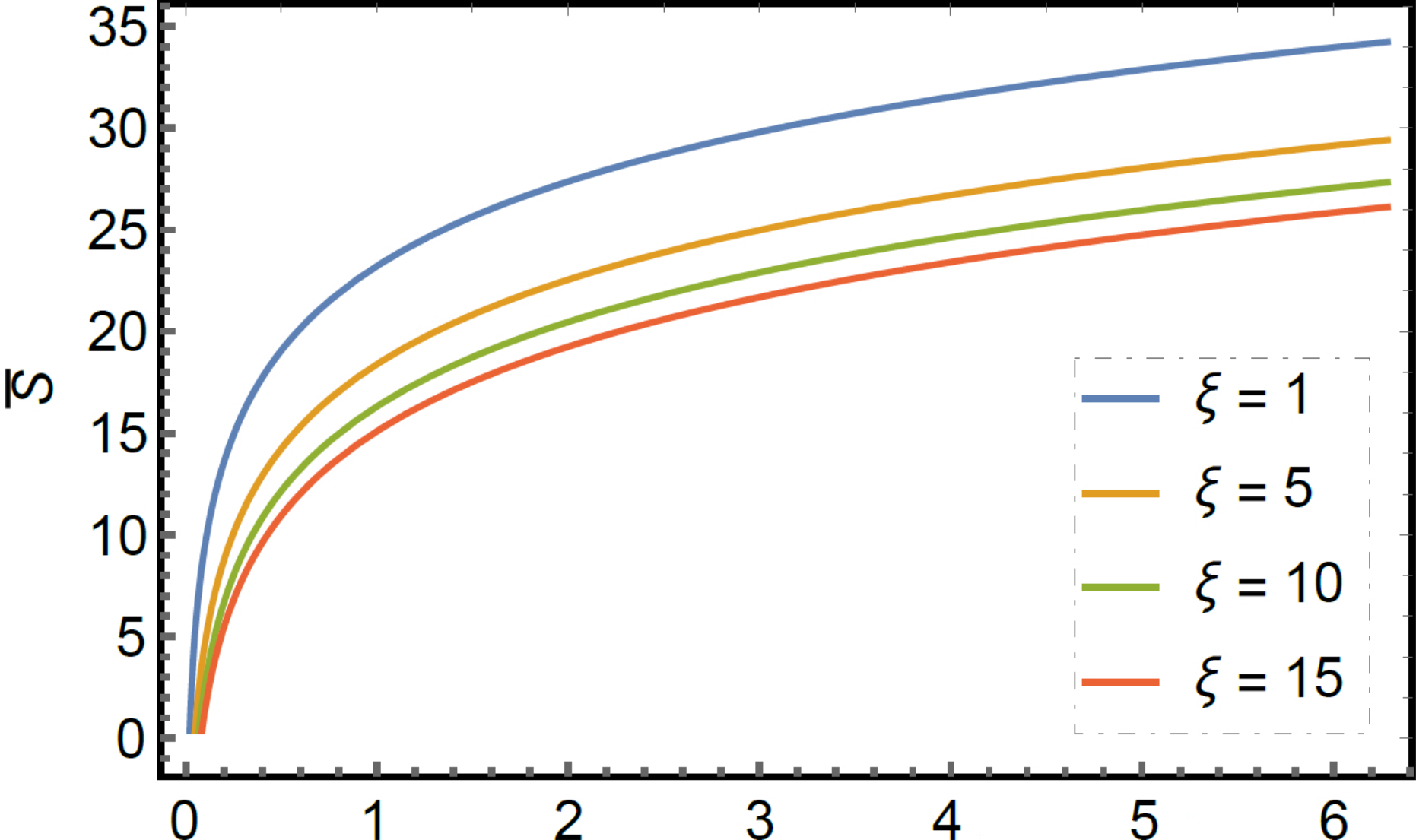}
\caption{The graphic shows the entropy $\bar{S}$ of neutral Dirac particles in the presence of electromagnetic field for different values of $\xi$ when the temperature $T$ increases for the relativistic case.}
\label{2}
\end{figure}

Here, we plot the graphics of the thermal quantities by temperature for different values of the parameter $\xi$ which are directly related to the external electric field. Namely, these values of $\xi$ are: $\xi=1$, $\xi=5$, $\xi=10$, $\xi=15$, which entail in a cyclotron frequency within the range $10^{20}<\omega_c<10^{22}$ Hz. In addition, it is worth mentioning that we consider the natural units ($\hbar=c=k_B=1$), and for the sake of simplicity, the unit rest mass ($m_0=1$) is taken into account. In particular, let us consider the relativistic case, which from Fig. \ref{1}, the Helmholtz free energy $\bar{F}(T)$ has a little increase when $T$ starts increasing. Nevertheless, as we can see in general in $\bar{F}(T)$, it decreases sharply for high values of $T$ and increases for high values of parameter $\xi$. As a matter of fact, this result shows that the equilibrium state is reached more slowly when $T$ decreases and $\xi$ increases. We see from Fig. \ref{2} that the entropy $\bar{S}(T)$ grows abruptly in the range $0<T<0.4$ and afterwards shows a smoother growth for large $T$, i.e., $\bar{S}(T)$ is a monotonically increasing function. Besides, the entropy grows more slowly with an increase of $\xi$, which implies that the equilibrium state is also reached more slowly with the decrease of $T$ and the increase of $\xi$.

\begin{figure}[ht]
\centering
\includegraphics[scale=0.581]{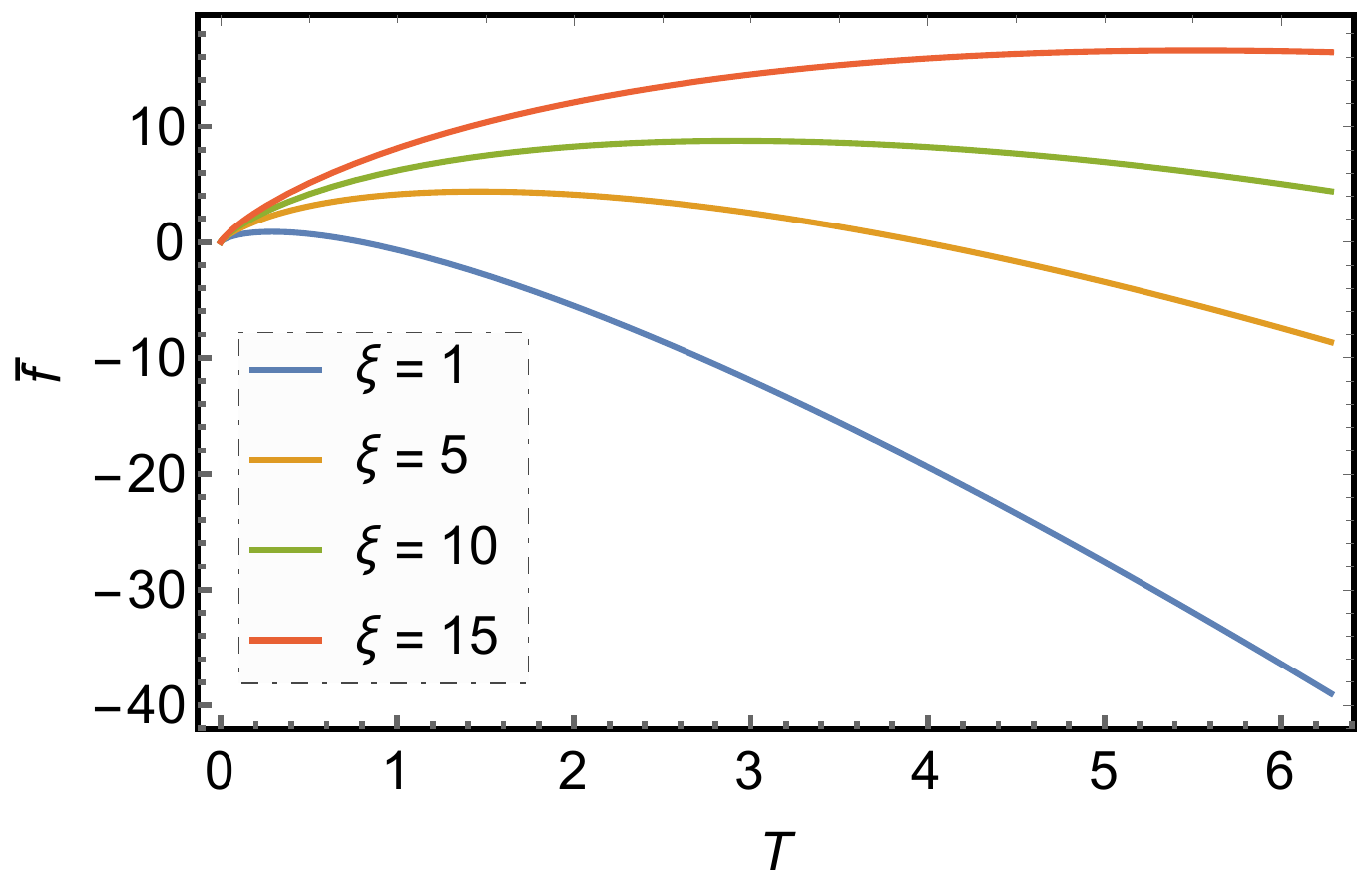}
\includegraphics[scale=0.43]{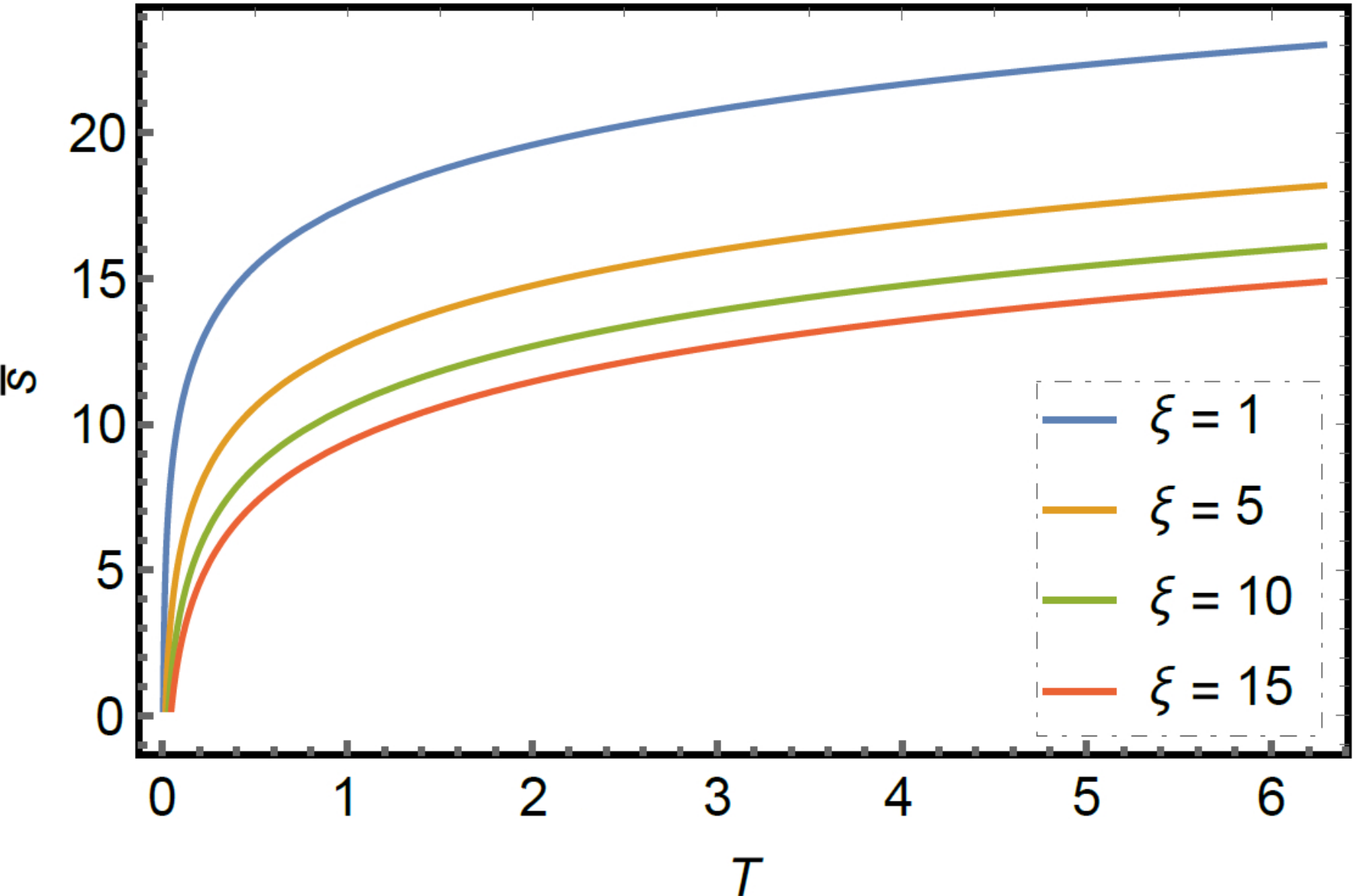}
\caption{The two graphs exhibit the behavior of Helmholtz free energy $\bar{f}$ and entropy $\bar{S}$ of non-relativistic neutral particles in the presence of an electromagnetic field for different values of $\xi$.}
\label{3}
\end{figure}

Now, let us focus on the comparison on the Helmholtz free energy $\bar{f}(T)$ and entropy $\bar{s}(T)$, which are displayed in Fig. \ref{3}, with the relativistic case. We verify that $\bar{f}(T)$ for having the behavior of decreasing, more temperature is required when $\xi$ grows. However, $\bar{s}(T)$ exhibit similar behavior in both cases when different values $\xi$ are regarded.

\begin{figure}[ht]
\centering
\includegraphics[scale=0.581]{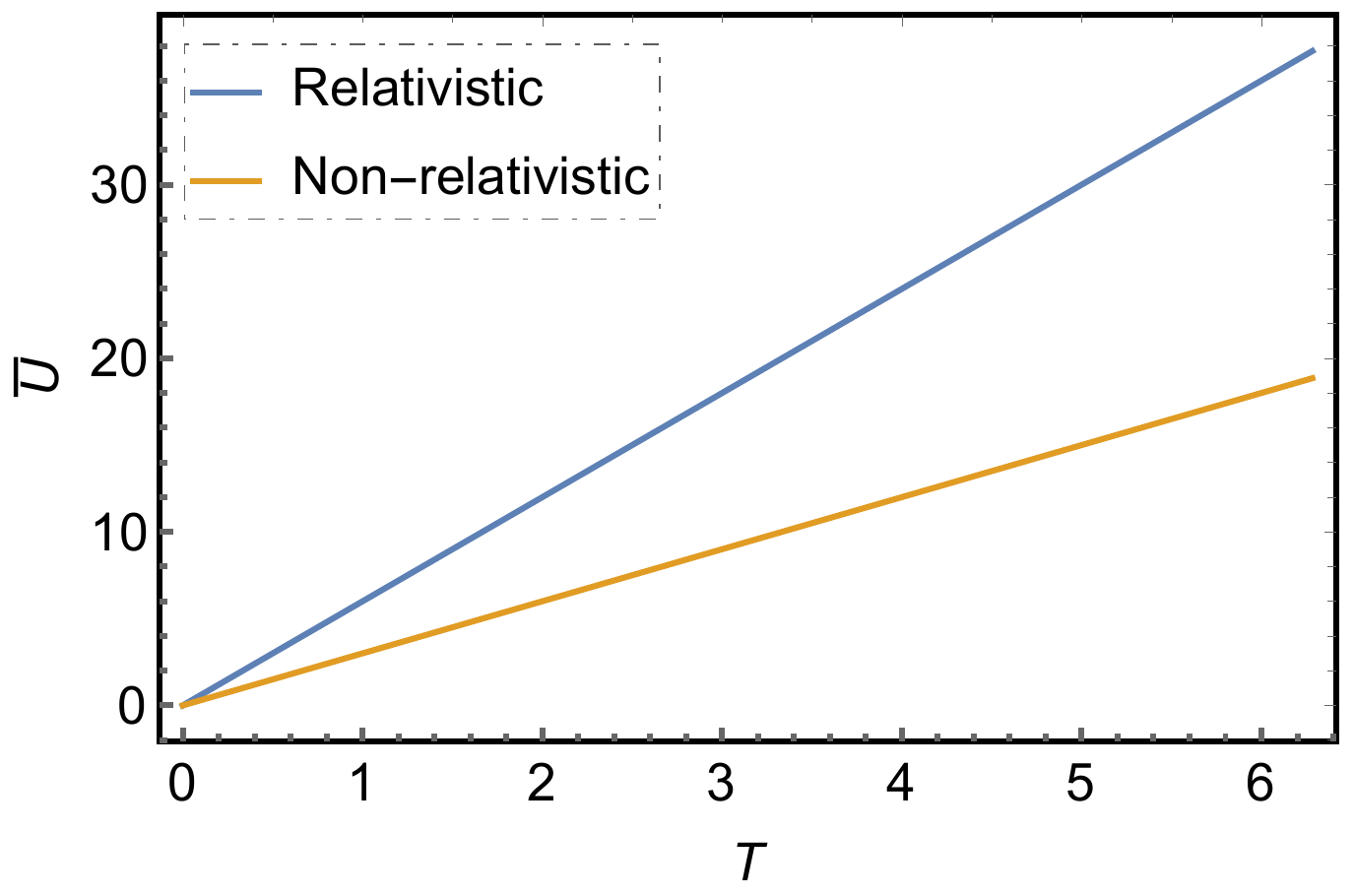}
\caption{The two lines represent the behavior of mean energy for both relativistic and non-relativistic cases.}
\label{4}
\end{figure}

In Fig. \ref{4}, we see that the relativistic and non-relativistic mean energies are displayed respectively. They increase entirely as a linear function when $T$ increases. Nevertheless, the relativistic mean energy increases faster than in the non-relativistic case, which is easily checked from Eqs. \eqref{properties1} and \eqref{properties2}. In particular, this result is due to the fact that in the relativistic case we have more energetic quantum system than in the non-relativistic case ($E>\varepsilon$). Moreover, the thermal variation of the ensemble of particles tends to increase until reaching the thermal equilibrium in both cases, and once the equilibrium is reached, the energy remains constant and the amount of free energy available to perform work is zero. Now, as we have $\bar{U}(T)>\bar{u}(T)$ and the mean energy can be written in the form $U(T)=F(T)+S(T)T$ \cite{Greiner}, implies that $\bar{F}(T)<\bar{f}(T)$ and $\bar{S}(T)>\bar{s}(T)$ for a given value of $T$, therefore, the behavior of the graphics in Figs.\ref{1} to \ref{4} are totally in agreement with the fundamentals of the statistical physics. From above, we verify that for a given value of $\xi$, the thermal equilibrium state is reached faster in the relativistic case than in the non-relativistic.

On the other hand, it is also seen that the heat capacity for the relativistic case is two times the value of the non-relativistic case as a result obtained from the calculations presented previously. According to the literature, when the mean energy and the specific heat capacity are twice of the value encountered in the non-relativistic case, we say that these cases satisfy the so-called \textit{Dulong-Petit} law \cite{Victor}. It is important to mention that this peculiar behavior is also observed in others relativistic quantum systems, such as in the Dirac oscillator \cite{Pacheco} and in the Aharonov-Bohm quantum ring \cite{Oli}. Last but not least, it is worth mentioning that, since the partition function in both cases does not depend on the magnetic field (Eq. \eqref{partition6} and Eq. \eqref{partition11}), there exists no influence on its thermodynamic functions.

\section{Conclusion\label{conclusion}}

As mentioned previously, this paper has the purpose of investigating the thermodynamic properties of neutral Dirac particles in the presence of an external electromagnetic field in contact with a heat bath. In this sense, we determine the thermodynamic properties for both relativistic and non-relativistic cases focusing on the high-temperature regime. In order to perform the calculations, we use the so-called \textit{Euler-MacLaurin} formula in order to obtain the canonical partition function of the system. Next, we determine the thermodynamic quantities of interest, namely, Helmholtz free energy, the mean energy, the entropy, and the heat capacity. 

Considering initially the relativistic case, we see that the Helmholtz free energy has a little increase when $T$ starts to increase, nevertheless, in general, it decreases sharply for high values of $T$ and increases for high values of parameter $\xi$ which is associated to the electric field. This result shows that the equilibrium state is reached more slowly with a decrease of $T$ and an increase of $\xi$. We see that the entropy function exhibits a monotonically increasing function. 

We verify that for both relativistic and non-relativistic cases, the mean energy turns out to behave as a linear function being the first one with a greater angular coefficient. Moreover, we verify that for different values of $\xi$, the thermal equilibrium state is reached more quickly in the relativistic case. On the other hand, it is also seen that the heat capacity regarding the relativistic case is twice the value presented in the non-relativistic case. As a matter of fact, in agreement with the literature, these results satisfy the famous \textit{Dulong-Petit} law. Finally, we verify that the external magnetic field has no influence on the behavior of thermodynamic quantities performed in this whole work.

\section*{Acknowledgments}

\hspace{0.5cm}The authors would like to thank the Conselho Nacional de Desenvolvimento Cient\'{\i}fico e Tecnol\'{o}gico (CNPq) for financial support.


\begin{thebibliography}{99}

\bibitem{Gaskell} D. R. Gaskell, D. E. Laughlin, {\it Introduction to the Thermodynamics of Materials}, 6nd ed., CRC press, 2018.

\bibitem{DeHoff} R. DeHoff, {\it Thermodynamics in Materials Science}, 2nd ed., CRC Press, 2006.

\bibitem{Tester} J. W. Tester, Michael Modell, {\it Thermodynamics and Its Applications}, 3rd Edition, Upper Saddle River, Prentice Hall, 1997.

\bibitem{Dolling} G. Dolling, R. A. Cowley, Proc. Phys. Soc. {\bf 88} (1966) 463.

\bibitem{Muhlschlegel} B. Muhlschlegel, D. J. Scalapino, R. Denton, Phys. Rev. B {\bf 6} (1972) 1767.

\bibitem{Coleman} B. D. Coleman, W. Noll, {\it The thermodynamics of elastic materials with heat conduction and viscosity. In The Foundations of Mechanics and Thermodynamics} (pp. 145-156), Springer, Berlin, Heidelberg (1974).

\bibitem{Eckert} J. Eckert, J. C. Holzer, C. E. Krill, W. L. Johnson, J. Mater. Res. {\bf 7} (1992) 1751-1761.

\bibitem{Foiles} S. M. Foiles, J. B. Adams, Phys. Rev. B {\bf 40} (1989) 5909.

\bibitem{Anthony} J. L. Anthony, E. J. Maginn, J. F. Brennecke, J. Phys. Chem. B {\bf 106} (2002) 7315-7320.

\bibitem{Wang} Y. Wang, Z. K. Liu, L. Q. Chen, Acta Mater. {\bf 52} (2004) 2665-2671.

\bibitem{Balandin} A. A. Balandin, Nat. Mater. {\bf 10} (2011) 569.

\bibitem{Mounet} N. Mounet, N. Marzari, Phys. Rev. B {\bf 71} (2005) 205214.

\bibitem{Shahil} K. M. Shahil, A. A. Balandin, Solid State Commun. {\bf 152} (2012) 1331-1340.

\bibitem{Pop} E. Pop, V. Varshney, A. K. Roy, MRS Bull. {\bf 37} (2012) 1273.

\bibitem{Alofi} A. Alofi, G. P. Srivastava, Phys. Rev. B {\bf 87} (2013) 115421.

\bibitem{Che} J. Che, T. Cagin, W. A. Goddard, Nanotechnology {\bf 11} (2000) 65.

\bibitem{Ruoff} R. S. Ruoff, D. C. Lorents, carbon {\bf 33} (1995) 925.

\bibitem{Boumali} A. Boumali, H. Hassanabadi, Eur. Phys. J. Plus {\bf 128} (2013) 124.

\bibitem{Hassanabadi} H. Hassanabadi, S. S. Hosseini, A. Boumali, S. Zarrinkamar, J. Math. Phys. {\bf 55}, 033502 (2014); H. Hassanabadi, S. Sargolzaeipor, B. H. Yazarloo, Few-Body Syst. {\bf 56} (2015) 115-124.

\bibitem{Pacheco} M. H. Pacheco, R. R. Landim, and C. A. S. Almeida, Phys. Lett. A {\bf 311} (2003) 93; M. H. Pacheco, R. V. Maluf, C. A. S. Almeida, R. R. Landim, Europhys. Lett. {\bf 108} (2014) 10005.

\bibitem{Wang2015} Z. Wang, Z. W. Long, C. Y. Long, W. Zhang, Adv. High Energy Phys. {\bf 2015}, (2015).

\bibitem{Arda} A. Arda, C. Tezcan, R. Sever, Few-Body Syst. {\bf 57} (2016) 93-101.

\bibitem{Groote} J. J. S. De Groote, J. E. M. Hornos, A. V. Chaplik, Phys. Rev. B {\bf 46} (1992) 12773.

\bibitem{Dong} S. H. Dong, M. Lozada‐Cassou, J. Yu, F. Jim\'enez‐\'Angeles, A. L. Rivera, Int. J. Quantum Chem. {\bf 107} (2007) 366-371.

\bibitem{Nammas} F. S. Nammas, A. S. Sandouqa, H. B. Ghassib, M. K. Al-Sugheir, Physica B: Condensed Matter {\bf 406} (2011) 4671-4677.

\bibitem{Oyewumi} K. J. Oyewumi, B. J. Falaye, C. A. Onate, O. J. Oluwadare, W. A. Yahya, Mol. Phys. {\bf 112} (2014) 127-141.

\bibitem{Arda2016} A. Arda, C. Tezcan, R. Sever, Eur. Phys. J. Plus {\bf 131} (2016) 323.

\bibitem{Oli} R. R. S. Oliveira, A. A. Ara\'ujo Filho, F. C. E. Lima, R. V. Maluf, C. A. S. Almeida, Thermodynamic properties of an Aharonov-Bohm quantum ring, arXiv:1812.08607, 2018 [quantph].

\bibitem{Hagen} C. R. Hagen, Phys. Rev. Lett. {\bf 64} (1990) 2347.

\bibitem{Mirza} B. Mirza, M. Zarei, Eur. Phys. J. C {\bf 32} (2004) 583–586.

\bibitem{Li} K. Li, J. Wang, Eur. Phys. J. C {\bf 50} (2007) 1007–1011.

\bibitem{Bakke} K. Bakke, C. Furtado, Ann. Phys. {\bf 336} (2013) 489-504.

\bibitem{Oliveira} R. R. S. Oliveira, R. V. Maluf, C. A. S. Almeida, Exact solutions of the Dirac oscillator under the influence of the Aharonov-Casher effect in the cosmic string background, (2018). arXiv: 1810.11149.

\bibitem{Silenko} A. Y. Silenko, Russ. Phys. J. {\bf 48} (2005) 788-792.

\bibitem{Khalilov} V. R. Khalilov, Theor. Math. Phys. {\bf 129} (2001) 1357-1368.

\bibitem{Lin} Q. G. Lin, Phys. Rev. A {\bf 81}, 012710 (2010); Q. G. Lin, Phys. Rev. A {\bf 72} (2005) 042103.

\bibitem{Azevedo} F. S. Azevedo, E. O. Silva, L. B. Castro, C. Filgueiras, D. Cogollo, Ann. Phys. {\bf 362} (2015) 196–207.

\bibitem{Bakke2010} K. Bakke, Phys. Lett. A {\bf 374} (2010) 4642–4646.

\bibitem{Bakke2008} K. Bakke, J. R. Nascimento, C. Furtado, Phys. Rev. D {\bf 78} (2008) 064012; K. Bakke, C. Furtado, J. R. Nascimento, Eur. Phys. J. C {\bf 60} (2009) 501.

\bibitem{Bakke2012} K. Bakke, Int. J. Theor. Phys. {\bf 51} (2012) 759–771; Eur. Phys. J. B {\bf 85} (2012) 354.

\bibitem{Moshinsky} M. Moshinsky, A. Szczepaniak, J. Phys. A, Math. Gen. {\bf 22} (1989) L817.

\bibitem{Martinez} R. P. Martinez-y-Romero, H. N. N\'unez-Y\'epez, A. L. Salas-Brito, Eur. J. Phys. {\bf 16} (1995) 135; J. Bentez, R. M. y Romero, H. N. N\'unez-Yépez, A. L. Salas-Brito, Phys. Rev. Lett. {\bf 64} (1990) 1643.

\bibitem{O} R. R. S. Oliveira, R. V. Maluf, C. A. S. Almeida, Ann. Phys. {\bf 400} (2019) 1-8.

\bibitem{H} H. Hassanabadi, M. Hosseinpour, Eur. Phys. J. C {\bf 76} (2016) 553.

\bibitem{Rubens} R. R. S. Oliveira, M. F. Sousa, Braz. J. Phys. {\bf 49} (2019) 315.

\bibitem{Greiner} W. Greiner, Ludwig Neise, Horst Stocker, D. Rischke, {\it Thermodynamics and statistical mechanics}, Springer, New York, 1995.

\bibitem{Victor} V. Santos, R. V. Maluf, C. A. S. Almeida, Ann. Phys. {\bf 349} (2014) 402-410.

\end{thebibliography}
\end{document}